\documentclass[reprint,prb,aps,showpacs,twocolumn]{revtex4-1}
\usepackage{epsfig,color}
\usepackage{graphicx}
\usepackage{amsmath}
\usepackage{amssymb}
\usepackage{float}
\restylefloat{table}

\begin{document}
\title{Single-electron shell occupation and effective $g$-factor\\ in few-electron nanowire quantum dots}

\author{M. P. Nowak}
\altaffiliation{Present address: QuTech and Kavli Institute of Nanoscience, Delft University of Technology, 2600 GA Delft, The Netherlands.}
\affiliation{AGH University of Science and Technology, Faculty of Physics and Applied Computer Science,
al. Mickiewicza 30, 30-059 Krak\'ow, Poland}

\author{B. Szafran}
\affiliation{AGH University of Science and Technology, Faculty of Physics and Applied Computer Science,
al. Mickiewicza 30, 30-059 Krak\'ow, Poland}
\date{\today}

\begin{abstract}
Nanowire double quantum dots occupied by an even number of electrons are investigated in the context of energy level structure revealed by electric dipole spin resonance measurements. We use numerically exact configuration interaction approach up to 6 electrons for systems tuned to Pauli spin blockade regime. We point out the differences between the spectra of systems with two and a greater number of electrons. For two-electrons the unequal length of the dots results in a different effective $g$-factors in the dots as observed by the recent experiments. For an increased number of electrons the $g$-factor difference between the dots appears already for symmetric systems and it is greatly amplified when the dots are of unequal length. We find that the energy splitting defining the resonant electric dipole spin frequency can be quite precisely described by the two electrons involved in the Pauli blockade with the lower-energy occupied states forming a frozen core.
\end{abstract}


\maketitle
\section{Introduction}
Few-electron gate-defined quantum dots\cite{hanson} are exploited for single spin manipulation that allows for realization of single-qubit quantum gates.\cite{loss} While the desired spin rotation involves a single spin as a carrier of quantum information, multielectron systems provide a feasible environment for readout of the controlled spin. Strong spin-orbit (SO) coupling that is present in InSb and InAs nanowires \cite{fasth, pfund} allows for electrical spin rotations\cite{nowack} that are performed by electric dipole spin resonance (EDSR)\cite{golovach} which excludes the need for introduction of local magnetic field gradients\cite{laird,koppens} or usage of hyperfine interaction.\cite{osika} The readout of the spin is realized via spin to charge conversion that relies on the Pauli spin blockade.\cite{ono} The single electron current $(1,0)\rightarrow(1,1)\rightarrow(2,0)\rightarrow(1,0)$ [the numbers in the brackets correspond to the number of electrons in a particular dot] is blocked at the transition from $(1,1)$ triplet to $(2,0)$ singlet. Rotation of one of the spins of electrons constituting the $(1,1)$ triplet unblocks the current which serves as a proof for the coherent spin control.
On the other hand strong SO coupling leads to unavoidable spin relaxation which results in a spontaneous lifting of the Pauli blockade when one of the $(1,1)$ triplets is close in energy to the $(2,0)$ singlet.\cite{nowak2014}

EDSR lifting of the current blockade is observed already for two electrons bound in the double dot, which indeed is the case for many of the experiments.\cite{nadj-perge,nadj2012,schroer,frolov} However some of the experimentally studied devices consist an even number $N$ of electrons greater than two.\cite{berg,petersson, stehlik} In this case the system is biased such the Pauli blockade is between $(N-1,1)$ and $(N,0)$ states. It is assumed that such configuration is equivalent to the two-electron system.\cite{rossella} This approximation resembles the well established concept in chemistry, that the valence electrons are responsible for creation of bonds and the rest in the deep levels can be treated as the frozen core.\cite{fc}
This assumption seems questionable for quantum dots in which the single-electron shells are separated by much smaller energies than for the Coulomb potential,
nevertheless this problem has not been discussed by a theoretical study.  The present work addresses this issue.

We find that for a system with $N>2$ all but two electrons form closed singlet shells. This is in accordance with predictions of Hubbard model,\cite{tasaki} that appear as a consequence of Mattis-Lieb theorem,\cite{lieb} and which states that the lowest energy states posses the lowest spin (S=0). As a consequence, in general, the low-energy spectra of multielectron double quantum dots in the $(N-1,1)$ configuration resemble the spectra of quantum dots in $(1,1)$ configuration and the states have similar total spins. We find that the $(N-1,1)$ spectra can be well recreated by a configuration interaction calculation in which one excludes the single-electron orbitals that form the singlet shells. This is analogous to the frozen core approximation regardless of the fact that there are no orbital shells in quantum-dots.

The main finding of the work is that though a general resemblance of $N>2$ and $N=2$ spectra is found, the occupation of excited single-electron orbitals in the $N>2$ case leads to lifting of the degeneracy of spin-zero states. This in turn is translated to different effective $g$-factors in the dots. Such differences have been observed in recent EDSR experiments on nanowire quantum dots\cite{nadj-perge,nadj2012,schroer,frolov,berg,petersson, stehlik} and have been related to the differences in the confinement as predicted by the study on self-organized quantum dots.\cite{pryor} Here we strictly connect the effective $g$-factors with the number of electrons in the system and the length of the dots. We find that unequal effective $g$-factors for $N=2$ appear only for an asymmetric system but for $N>2$ they are observed already for the dots of the same lengths.

\section{Theory}
In the present work we follow the common approach\cite{oned} that treats the nanowire quantum dots as quasi-one dimensional. The N-electron system is described by the Hamiltonian,
\begin{equation}
H=\sum_{i=1}^N h_i + \sum_{i=1,j=i+1}^N \frac{\sqrt{\pi/2}}{4\pi\varepsilon_0\varepsilon \ell}\mathrm{erfcx}\left[\frac{|x_1-x_2|}{\sqrt{2}\ell}\right].
\label{hne}
\end{equation}
The form of the Coulomb interaction term results from the assumption that the electrons are localized in the ground state of the lateral quantization along the nanowire cross section with the  wavefunction of a Gaussian shape with $\psi(y,z)=(\pi^{1/2}\ell)^{-1}\exp\{-(y^2+z^2)/(2\ell^2)\}$. The integration of three-dimensional Coulomb interaction term $H_C= \sum_{i=1,j=i+1}^N \frac{e^2}{4 \pi \varepsilon \varepsilon_0} \frac{1}{|r_i-r_j|}$ leads\cite{bednarek} to the operator including $\mathrm{erfcx}(x)=\exp(x^2)\mathrm{erfc}(x)$ which is the exponentially scaled complementary error function.\cite{book}

To obtain N-electron spin-orbitals we diagonalize the Hamiltonian (\ref{hne}) in a basis of Slater determinants consisting of single-electron spin-orbitals.
\begin{equation}
\Psi(\nu_1,\nu_2,...,\nu_N)=\sum_{i=1}^M c_i \emph{A} \{\psi_{i_1}(\nu_1)\psi_{i_2}(\nu_2)...\psi_{i_N}(\nu_N)\},
\end{equation}
where $\nu_i=(x_i,\sigma^i)$ corresponds to the orbital and spin coordinates, $\emph{A}$ is the antisymmetrization operator and $c_i$ is obtained by the diagonalization. We use $M=20$ single-electron spin-orbitals which provides accuracy better than $0.1\;\mu$eV.

The single-electron orbitals $\psi(\nu)$ are described by the Hamiltonian,
\begin{equation}
h = \frac{\hbar^2 k_x^2}{2m^*} + V(x) - \alpha \sigma_y k_x + \frac{1}{2}\mu_B gB\sigma_x,
\label{1eham}
\end{equation}
where $H_{SO 1D}=-\alpha\sigma_yk_x$ corresponds to Rashba SO coupling\cite{rashba} resulting from $H_{SO}=\alpha(\sigma_xk_y-\sigma_yk_x)$ Hamiltonian averaged in the $y$-direction.
$V(x)$ describes the potential profile of the double dot,
\begin{equation}
V(x) = \left\{
  \begin{array}{l l}
    V_b & \quad x<-w/2\;\textrm{and}\;x>-(l_1+w/2)\\
    V_i & \quad |x|<w/2\\
    0 & \quad x>w/2\;\textrm{and}\;x<l_2+w/2
  \end{array} \right.
\end{equation}
where $l_1$ and $l_2$ determine the length of each dot, $w$ is the interdot barrier width, $V_i$ is the barrier height and $V_b$ is the bias potential applied to the bottom of the left dot. We assume a $w=20$ nm thin and $V_i=200$ meV high interdot barrier. The computational box ends at the edges of the defined potential and the magnetic field is applied along the nanowire axis. The single-electron eigenstates are obtained by exact diagonalization of Hamiltonian (\ref{1eham}) on a mesh of 201 points with $\Delta x = 1.095$ nm.

We adopt parameters corresponding to InSb nanowires, i.e. $m^*=0.014$, $\varepsilon=16.5$, $g=-51$ and $\alpha=30$ meVnm which corresponds to spin-orbit length $l_{so}=\hbar/(m^*\alpha)=182$ nm comparable to the value measured experimentally in Ref. \onlinecite{nadj2012}. We take $\ell=20$ nm.

\section{Results}
\subsection{Two-electron quantum dot}
\begin{figure}[ht!]
\epsfxsize=75mm
                \epsfbox[20 54 567 800] {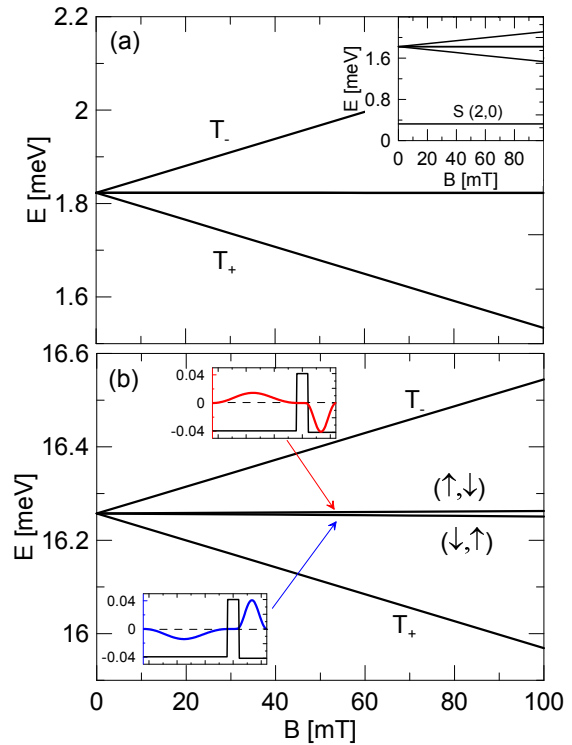}
                 \caption{Energy spectra of two-electron double dot system. (a) Symmetric system with the dots width equal to 100 nm. (b) Asymmetric configuration: width of the left dot is 150 nm and the width of the right one is 50 nm. Inset in (a) shows the energy spectrum including the ground state (2,0) singlet. The insets in (b) depict the spin densities with the color curves presented in the arbitrary units. The black contours show the confinement potential.}
 \label{spectr2e}
\end{figure}

Let us first consider a symmetric system of two quantum dots of lengths $l_1=l_2=100$ nm. We set the bottom of the left dot to $V_b=-3.8$ meV. The bias results in the energy level configuration such that $(2,0)$ singlet\cite{comment} is the ground state and the lowest-energy excited states are $(1,1)$ states with different spin polarizations. This configuration is necessary for observation of spin Pauli blockade. The inset to Fig. \ref{spectr2e}(a) shows the lowest part of the energy spectrum. The ground state singlet of (2,0) occupation has mean value of $<S^2>$ operator 0.12 $\hbar^2/4$.  Figure \ref{spectr2e}(a) presents energy levels of (1,1) states. The two Zeeman split energy levels correspond to a spin-positive triplet $T_+$ ($<S^2>=1.98\;\hbar^2/4$) with spins oriented approximately along the magnetic field and to a spin-negative triplet $T_-$ ($<S^2>=1.97\;\hbar^2/4$) with spins oriented against the magnetic field. The horizontal curve corresponds to a degenerate energy level of a singlet (S) and a triplet ($T_0$) states with zero spin projection along the direction of the magnetic field. The degeneracy results from the negligible overlap between the adjacent electrons and hence nearly zero exchange interaction. The mean values of $<S^2>$ operator for these states are: 1.04, 1.01 $[\hbar^2/4]$.

\begin{figure}[ht!]
\epsfxsize=70mm
                \epsfbox[19 200 570 640] {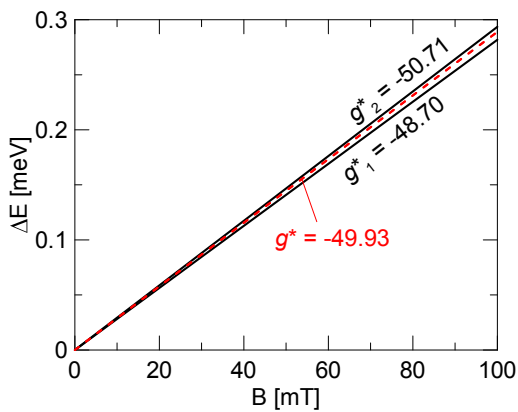}
                 \caption{Energy difference (the red-dashed curve) between the energy level of $T_+$ state and the degenerated energy level from Fig. \ref{spectr2e}(a) or two levels that correspond to states with opposite spin configurations from Fig. \ref{spectr2e}(b) (black curves).}
 \label{2ediffen}
\end{figure}

In EDSR experiments the spin rotations are performed from one of the non-zero spin triplets: $T_+$ or $T_-$.\cite{nowak2014} When a resonance to a state with zero spin component along the magnetic field occurs the blockade is lifted. The experimentally measured resonances exhibit a linear dependence of the driving frequency on the magnetic field, equal to (considering $T_+$ as the initial state) $\omega=[E(S)-E(T_+)]/\hbar$. The corresponding energy $\omega\hbar$ is plotted in Fig. \ref{2ediffen} with the red-dashed curve.

\begin{figure*}[ht!]
\epsfxsize=170mm
                \epsfbox[25 339 570 515] {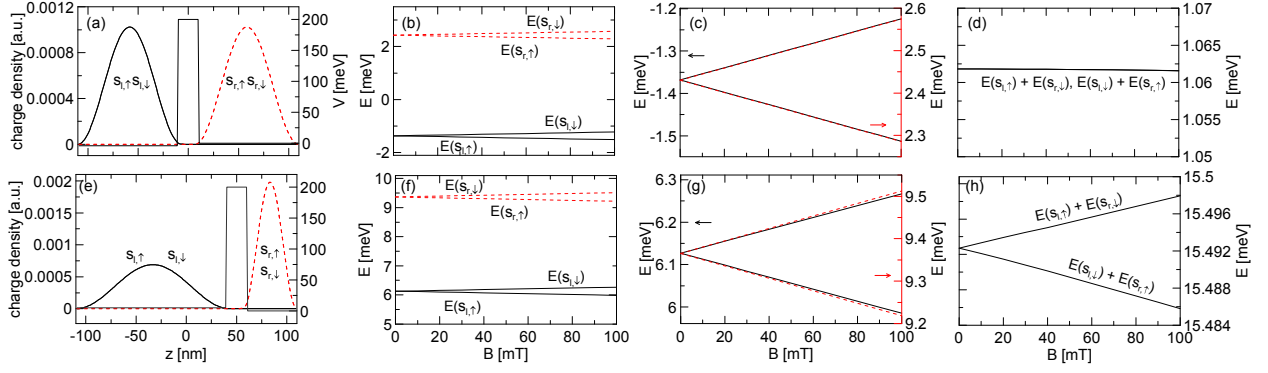}
                 \caption{(a,e) Single electron charge densities and the potential profile. (b,f) Energy spectrum -- the colors of the curves corresponds to the states from (a,e). (c,g) Energy levels from (b,f) shifted to compare the Zeeman splittings. (d,h) Sum of the single-electron energies as they enter into configuration interaction approach.}
 \label{spectr1e}
\end{figure*}

Let us now consider the case in which the dots are of unequal length -- $l_1=150$ nm and $l_2=50$ nm -- but we keep the $g$-factor constant along the structure. Energy levels of states in which electrons occupy adjacent dots are presented in Fig. \ref{spectr2e}(b). To keep the energy separation between (2,0) singlet and four (1,1) states 1.5 meV at $B=0$ as in the case of symmetric system we set $V_{bias}=5.07$ meV in the left dot. The striking difference between the spectra of Fig. \ref{spectr2e}(b) as compared to the symmetric case of Fig. \ref{spectr2e}(a) is that for the non-zero magnetic field the degeneracy of horizontal energy levels is lifted. The spin densities of the states that correspond to these energy levels calculated as $\sigma^j_x(x)=\sum_{i=1}^N\langle\Psi_j(\nu_1,\nu_2,...,\nu_N)|\sigma^i_{x}\delta_{x_i,x}|\Psi_j(\nu_1,\nu_2,...,\nu_N)\rangle$ are depicted in the insets to Fig. \ref{spectr2e}(b). We observe that in the state with lower energy the spin in the left dot is oriented against the magnetic field, while in the right dot it is oriented along the magnetic field. Further on we will address to this state as to $(\downarrow,\uparrow)$. The state $(\uparrow,\downarrow)$ with an increasing energy in $B$ has an opposite spin configuration. The $(\downarrow,\uparrow)$, $(\uparrow,\downarrow)$ states have zero total spins along the $x$-direction therefore the EDSR transitions to this states lift the Pauli blockade and such transition are visible as a resonance lines in EDSR spectra. We calculate corresponding energy differences $\Delta E_1 =\omega_1 \hbar=[E(\downarrow,\uparrow)-E(T_+)]$, $\Delta E_2 =\omega_2 \hbar=[E(\uparrow,\downarrow)-E(T_+)]$ and plot them in Fig. \ref{2ediffen} with black solid lines. We note that this arrangement of resonance lines is present in every EDSR map registered experimentally [see Refs. \onlinecite{nadj-perge,nadj2012,berg,schroer,frolov,petersson, stehlik}] and attributed to different $g$-factors in the dots. Here it is obtained for a constant $g$ along the structure.

The slopes of the curves in Fig. \ref{2ediffen} are connected to {\it effective} $g$-factor. For symmetric system we calculate $g^*=\frac{E(S)-E(T_+)}{\mu_0 B}$ equal to $g^*=-49.93$ for $B=100$ mT. In the case of asymmetric dots the effective $g$-factors are $g_1^*=-48.70$, $g_2^*=-50.71$.

To explain the impact of the dots width on the energy spectra and effective $g$-factors let us inspect the single electron spin-orbitals that constitute the two-electron orbitals. Figures \ref{spectr1e} (a,e) show the charge densities of the single-electron states. The densities correspond to the ground states of orbital quantization of each dot. The black curves correspond to $s_{l,\uparrow}$ and $s_{l,\downarrow}$ states [the main letter denotes the orbital excitation, $(l,r)$ denotes the dot in which the electron is localized and the arrows correspond to the average spin polarization direction]. The red-dashed curve shows the charge densities of higher energy states $s_{r,\uparrow}$ and $s_{r,\downarrow}$ in which the electron occupies the right dot. We extract the squared absolute values of coefficients -- $|c_i|^2$ -- for each of the Slater determinant that is used in the configuration interaction approach. For the symmetric case we get 0.806 for the determinant consisting of $\{s_{l,\uparrow},s_{r,\downarrow}\}$ single-electron orbitals and 0.194 for consisting $\{s_{l,\downarrow},s_{r,\uparrow}\}$ orbitals for one of the states from the degenerate pair of spin zero two-electron states (the coefficients for the second state are reversed). For the asymmetric case we get 0.992 for $\{s_{l,\downarrow},s_{r,\uparrow}\}$ for the $(\downarrow,\uparrow)$ state and 0.996 for $\{s_{l,\uparrow},s_{r,\downarrow}\}$ for the $(\uparrow,\downarrow)$ state. The lack of the admixture of other Slater determinants is due to small size of the dots which results in a considerable kinetic energy separation of the single-electron orbitals. The energy spectra displayed in Fig. \ref{spectr1e}(b,f) show the Zeeman splittings of the single-electron energy levels. If we overlay the energy levels of the states in which the electron is localized in the left and right dot (solid black and red-dashed curves in Fig. \ref{spectr1e}(b,f) respectively) we observe that they are exactly the same [Fig. \ref{spectr1e}(c)] when the dots are of identical length but they differ when the dots are of unequal length [Fig. \ref{spectr1e}(g)]. Now we sum the single-electron energies accordingly to the way the states enter the configuration-interaction calculation, i.e. the $(\downarrow,\uparrow)$ state corresponds to the occupation of $s_{l,\downarrow}$, $s_{r,\uparrow}$ single-electron orbitals with energies $E(s_{l,\downarrow})$ and $E(r_{r,\uparrow})$. The $(\uparrow,\downarrow)$ corresponds to the occupation of $s_{l,\uparrow}$ , $s_{r,\downarrow}$ single-electron orbitals with energies $E(s_{l,\uparrow})$ and $E(s_{r,\downarrow})$. The obtained sums are plotted in Fig. \ref{spectr1e}(d) for the symmetric and in Fig. \ref{spectr1e}(h) for the asymmetric case. We find that the degeneracy is lifted due to different Zeeman splittings of single-electron energy levels of electrons confined in dots of different length.

\begin{figure}[ht!]
\epsfxsize=60mm
                \epsfbox[107 18 503 823] {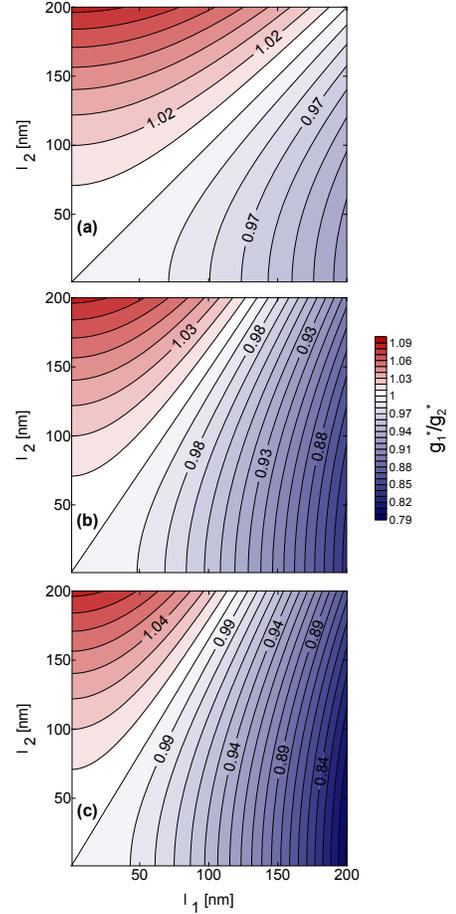}
                 \caption{$g_1^*/g_2^*$ versus the length of the dots for (a) two-electrons, (b) four-electrons and (c) six-electrons.}
 \label{g1g2}
\end{figure}

In a single one-dimensional quantum dot SO interaction impacts the Zeeman splittings according to $E_z=g\mu_B B\lambda_i$ where\cite{nowaksp}
\begin{equation}
\lambda_i=\int|\Psi_i(x)|^2\cos(2\alpha m^*x/\hbar^2) dx.
\label{integral}
\end{equation}
Due to a high interdot barrier we can effectively treat the considered system as two separate dots. For a quantum dot in a form of an infinite quantum well the term $\lambda_i$ that controls the strength of the Zeeman splitting is
\begin{equation}
\lambda_1(l)=\frac{\hbar^6\pi^2\sin(l\alpha m^*/\hbar^2)}{\alpha m^* l (\pi^2 \hbar^4-\alpha^2{m^*}^2l^2)},
\end{equation}
for $\Psi_i(x)$ in a form of $s$-like orbital.
$\lambda_i$ changes from 1 for narrow quantum dots to 0 in the limit of infinite dot length. Accordingly, the Zeeman splittings are the strongest (as strong as in the absence of SO coupling) for a narrow quantum dot and become weaker if the length of the dot is increased.
The $g^*$-factors calculated from $\Delta E_1 =\omega_1 \hbar=[E(\downarrow,\uparrow)-E(T_+)]$ and $\Delta E_2 =\omega_2 \hbar=[E(\uparrow,\downarrow)-E(T_+)]$ depend on the Zeeman splitting of the single electron energy levels as follows [taking $E_0$ as the orbital excitation energy of $(\downarrow,\uparrow)$,$(\uparrow,\downarrow)$ and $T_+$ states]: $\Delta E_1 = E_0 + E(s_{l,\downarrow})+E(s_{r,\uparrow})-E_0-E(s_{l,\uparrow})-E(s_{r,\uparrow})=E(s_{l,\downarrow})-E(s_{l,\uparrow})=g\mu_BB\lambda_1(l_1)$
 and $\Delta E_2 = E_0+ E(s_{l,\uparrow})+E(s_{r,\downarrow})-E_0-E(s_{l,\uparrow})-E(s_{r,\uparrow})=E(s_{r,\downarrow})-E(s_{r,\uparrow})=g\mu_BB\lambda_1(l_2)$. Therefore $g_1^*=g\lambda_1(l_1)$ and $g_2^*=g\lambda_1(l_2)$. We plot the ratio $g_1^*/g_2^*=\lambda_1(l_1)/\lambda_1(l_2)$ in Fig. \ref{g1g2}(a). For $l_1=150$ and $l_2=50$ we obtain $g_1^*/g_2^*=0.960$ which matches well the value obtained in the exact calculation of Fig. \ref{2ediffen}, $g_1^*/g_2^*=0.961$

It should be noted here that the effect of the SO interaction on the strength of the Zeeman splittings is influenced also by the orientation of the magnetic field.\cite{nowaksp} For the magnetic field vector forming a $\phi$ angle with the nanowire axis the splitting becomes $E_z=g\mu_B B \sqrt{1-(1-\lambda_i^2)\cos^2\phi}$, i.e. the $g^*$ values obtained for the magnetic field oriented perpendicular to the nanowire axis approach the bulk $g$-factor value.

\subsection{Four- and six-electron case}
\begin{figure}[ht!]
\epsfxsize=70mm
                \epsfbox[40 20 550 820] {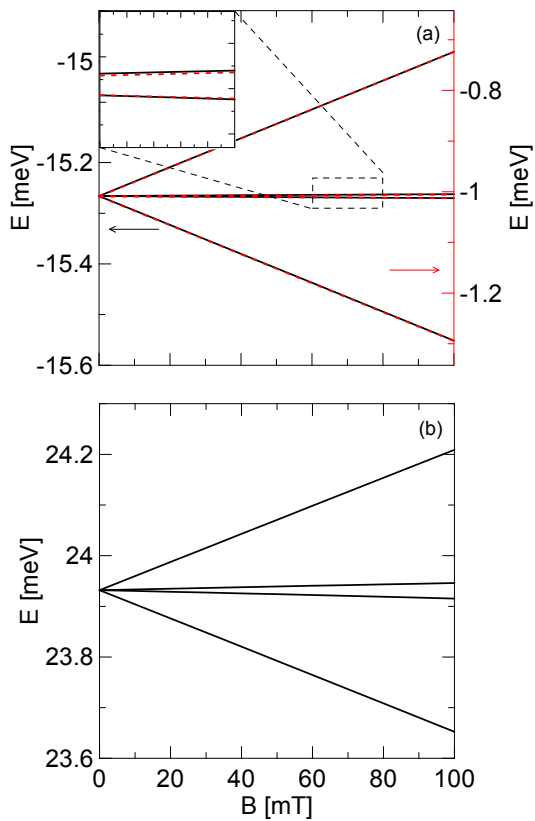}
                 \caption{(a) Energy levels of four-electron symmetric double dot in (3,1) occupation regime depicted with black-solid curves. The red-dashed curves are energy levels of a two-electron system with excluded $s_{l,\uparrow}$ and $s_{l,\downarrow}$ single-electron orbitals that form a singlet closed shell. (b) Energy spectrum for asymmetric system with $l_1=150$ nm and $l_2=50$ nm.}
 \label{spectr4e}
\end{figure}

Figure \ref{spectr4e}(a) with black solid curves presents energy levels of four-electron symmetric system with $V_{b}=-14.21$ meV. The levels correspond to the states with (3,1) occupation. The plot omits the ground state with (4,0) occupation that is $1.5$ meV lower in energy with respect to presented energy levels for $B=0$. The mean value of $<S^2>$ operator for the following (3,1) states are: 1.97, 1.04, 1.03, 1.97 $[\hbar^2/4]$. These values are close to the ones obtained for the two electron system. The absence of total spins of two electrons shows that two electrons form a singlet state with zero total spin.

\begin{figure}[ht!]
\epsfxsize=85mm
                \epsfbox[21 290 570 550] {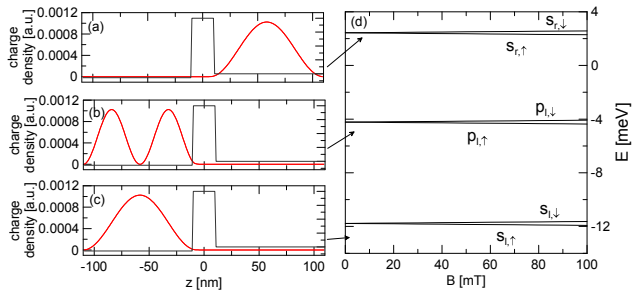}
                 \caption{(a,b,c) Charge densities of single-electron states and the potential profile for symmetric quantum dots with $V_b=-14.21$ meV. (d) Single electron energy spectrum.}
 \label{1efor4e}
\end{figure}

Let us extract the coefficients for each Slater determinant that is used to create configuration interaction basis. For the subsequent states whose energy levels are depicted in Fig. \ref{spectr4e}(a) the only non-zero (and nearly equal to unity, the other coefficients are less than 0.006) are coefficients for Slater determinants consisting of following single electron orbitals: $\{s_{l,\uparrow}s_{l,\downarrow}p_{l,\uparrow}s_{r,\uparrow}\}$, $\{s_{l,\uparrow}s_{l,\downarrow}p_{l,\downarrow}s_{r,\uparrow}\}$, $\{s_{l,\uparrow}s_{l,\downarrow}p_{l,\uparrow}s_{r,\downarrow}\}$ and $\{s_{l,\uparrow}s_{l,\downarrow}p_{l,\downarrow}s_{r,\downarrow}\}$ respectively. The corresponding orbitals are depicted in Fig. \ref{1efor4e}(a,b,c). Let us assume now that two electrons of the four-electron system form a singlet closed-shell that does not impact the spin properties of the two remaining electrons and thus they can be separated away: we exclude from the configuration interaction basis the $s_{l,\uparrow},s_{l,\downarrow}$ orbitals and limit number of electrons in the calculation to two.  The obtained energy levels are depicted with the red-dashed curves in Fig. \ref{spectr4e}(a). Besides the shift between the energy levels obtained in full four-electron and two-electron calculation with restricted basis the spectra perfectly match.

The total spins and the Zeeman splittings in the four-electron energy spectra resemble the ones obtained for two electrons. However, the striking feature of spectrum of Fig. \ref{spectr4e}(a) is that the two horizontal energy levels become separate in non-zero magnetic field already for a symmetric system. Let us invoke the two-electron approximation with the restricted basis to explain this observation. The $s_{l,\uparrow},s_{l,\downarrow}$ orbitals are occupied by two spin-opposite electrons that form the singlet state and are separated from the basis. The two energy levels that slightly split in the magnetic field are constructed from an $p$-like orbital of electron localized in the left dot ($p_{l,\uparrow}, p_{l,\downarrow}$) [see Fig. \ref{1efor4e}(b)] and an $s$-like orbital formed by an electron localized in the right dot ($s_{r,\uparrow},s_{r,\downarrow}$) [see Fig. \ref{1efor4e}(a)]. The single-electron energy levels are depicted in Fig. \ref{1efor4e}(d). We observe that the Zeeman splittings between energy levels of $s$-states differ from  the ones for $p$-orbitals: 0.282 meV compared to 0.294 meV. Here the dots are symmetric so it its the shape of $\psi_i(x)$ that is changed. We integrate Eq. \ref{integral} for an $p$-like orbital and obtain,
\begin{equation}
\lambda_2(l)=\frac{4 \hbar^6 \pi^2\sin(l\alpha m^*/\hbar^2)}{\alpha m^* l (4\pi^2\hbar^4-\alpha^2{m^*}^2l^2)}.
\end{equation}
The effective $g$-factors are obtained from the energy splittings analogically as in the two-electron case: $\Delta E_1=E_0+E(p_{l,\downarrow})+E(s_{r,\uparrow})-E_0-E(p_{l,\uparrow})-E(s_{r,\uparrow})=E(p_{l,\downarrow})-E(p_{l,\uparrow})=g\mu_BB\lambda_2(l_1)$ and $\Delta E_2=E_0+E(p_{l,\uparrow})+E(s_{l,\downarrow})-E_0-E(p_{l,\uparrow})-E(s_{r,\uparrow})=E(s_{r,\downarrow})-E(s_{r,\uparrow})=g\mu_BB\lambda_1(l_2)$. As a result the two states constructed from $\{p_{l,\downarrow},s_{r,\uparrow}\}$ and $\{p_{l,\downarrow},s_{r,\uparrow}\}$ single-electron orbitals have different energies at $B\ne0$ for $l_1=l_2$.

Figure \ref{g1g2}(b) presents $g_1^*/g_2^*=\lambda_2(l_1)/\lambda_1(l_2)$. The plot suggests that the $g$-factor ratio can be altered significantly as compared to the two-electron case for an asymmetric system. Namely, if one makes the dot that is occupied by three electrons longer one can amplify the ratio of the $g$-factors in the dots greater than elongating the dot with a single electron. The energy spectra for asymmetric system with $l_1=150$ nm and $l_2=50$ nm are presented in Fig. \ref{spectr4e}(b). The splitting between the central lines is visibly increased as compared to the symmetric case of Fig. \ref{spectr4e}(a). We calculate $g_1^*/g_2^*=0.896$ which is close to the value obtained in analytical calculation from Fig. \ref{g1g2}(b) equal to 0.910.

\begin{figure}[ht!]
\epsfxsize=70mm
                \epsfbox[48 32 540 820] {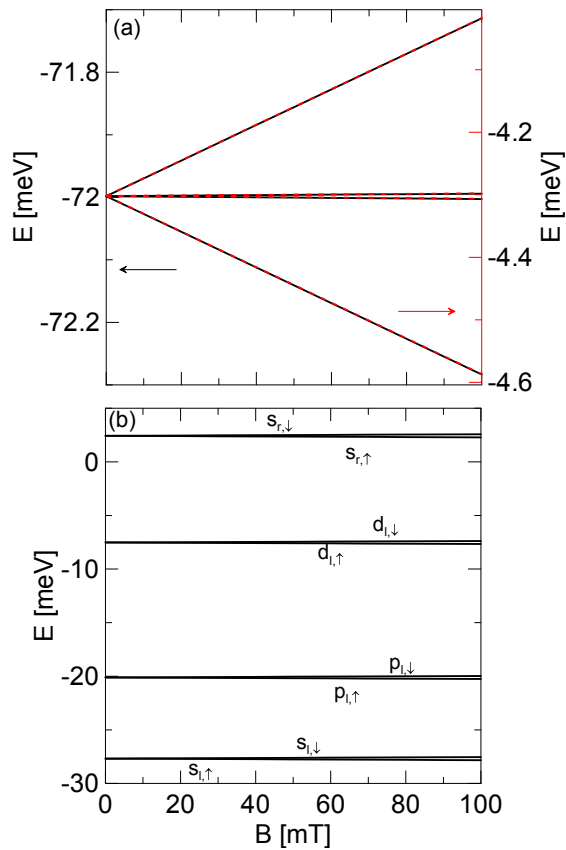}
                 \caption{(a) Six-electron energy spectrum for symmetric system with $V_b=-30.125$ meV. Black curves correspond to the exact six-electron calculation. Red-dashed curves are obtained in two-electron calculation with the basis excluding four lowest energy single-electron spin-orbitals. (b) Single electron energy spectrum.}
 \label{spectr6e}
\end{figure}

Figure \ref{spectr6e}(a) presents the energy spectrum of (5,1) states for six-electron double dot system for the bias $V_b=-30.125$ meV. The energy level structure resembles the spectrum of four-electron system depicted in Fig. \ref{spectr4e}(a). We again obtain the splitting of the central lines already for a symmetric system. For $B=100$ mT is 7.2 $\mu$eV for four electrons while for six electrons we get 8.6$\mu$eV. Also the total spins of (5,1) states are similar: 1.97, 1.04, 1.03, 1.97 $[\hbar^2/4]$. The coefficients for Slater determinants extracted from the configuration interaction calculation show that mainly a single determinant (with the square of absolute value equal to 0.987) describes each of the discussed six-electron state. The single electron states that constitute the determinants are: $\{s_{l,\uparrow},s_{l,\downarrow},p_{l,\uparrow},p_{l,\downarrow},d_{l,\uparrow},s_{r,\uparrow}\}$,
$\{s_{l,\uparrow},s_{l,\downarrow},p_{l,\uparrow},p_{l,\downarrow},d_{l,\downarrow},s_{r,\uparrow}\}$,
$\{s_{l,\uparrow},s_{l,\downarrow},p_{l,\uparrow},p_{l,\downarrow},d_{l,\uparrow},s_{r,\downarrow}\}$,
$\{s_{l,\uparrow},s_{l,\downarrow},p_{l,\uparrow},p_{l,\downarrow},d_{l,\downarrow},s_{r,\downarrow}\}$ for the following (5,1) states from Fig. \ref{spectr6e}(a). The single-electron energy levels are depicted in Fig. \ref{spectr6e}(b). The determinants correspond to the occupation of single-electron orbitals in which two pairs of electrons occupy closed singlet shells: two electrons occupy spin opposite $s$-like orbitals and the next pair occupies two spin opposite $p$-like orbitals. The two remaining electrons occupy an $d$-like orbital in the left dot and an $s$-like orbital in the right dot.  The calculated spectrum for the basis excluding the four single electron states that form the two singlet shells is presented in Fig. \ref{spectr6e}(a) with the red-dashed curves. The spectra obtained in the exact calculation and in the restricted basis agree.

For six electrons we calculate the ratio $g_1^*/g_2^*=\lambda_3(l_1)/\lambda_1(l_2)$ where
\begin{equation}
\lambda_3(l)=\frac{\hbar^6\sin(l\alpha m^*/\hbar^2)(9\pi^2\hbar^4-2\alpha^2{m^*}^2l^2)}{\alpha m^* l (9\pi^2 \hbar^4-\alpha^2{m^*}^2l^2)},
\end{equation}
is determined from integration of Eq. \ref{integral} with an $d$-like orbital and plot it in Fig. \ref{g1g2}(c). We see that it is similar to the four-electron case of Fig. \ref{g1g2}(b) and is strongly altered as compared to the $N=2$ case.

\subsection{Comparison with the experiments}

Our work shows that increasing the number of electrons results in an amplification of the difference between effective $g$-factors in the dots. Table I shows $g_1^*/g_2^*$ values taken from the experimental works. It is clearly seen that the studies that considered $N>2$ electrons indeed measured ratios that deviate more from 1 as compared to the $N=2$ cases. The actual experimental values could be affected by a number of effects omitted in the present modeling: the detailed structure of the confinement potential, or they can be impacted by non-zero exchange interaction. \cite{nowakharm} Nevertheless the tendency drawn by these data is clear and agrees with the result of the present study.

\begin{table}[H]
\centering
\begin{tabular}{|c|c|c|}
  \noalign{\hrule height 0.7pt}
  Reference No. & Number of electrons & $g_1^*/g_2^*$ \\
  \noalign{\hrule height 0.7pt}
  \onlinecite{nadj-perge} & N=2 & 0.967\\\hline
  \onlinecite{schroer} & N=2 & 0.923\\\hline
  \onlinecite{nadj2012} & N=2 & 0.922\\\hline
  \onlinecite{berg} & $N>2$ & 0.750\\\hline
  \onlinecite{petersson} & $N>2$ & 0.760\\\hline
  \onlinecite{stehlik} & $N=6$ & 0.872\\\hline
\end{tabular}
 \label{t1}
  \caption{$g_1^*/g_2^*$ ratio given in the experimental works versus the number of electrons.}
\end{table}

\section{Summary and conclusions}
We investigated nanowire double quantum dots occupied by an even number of electrons tuned to the Pauli spin blockade regime. By the exact configuration interaction study we found that in a system with an even, larger than two, number of electrons all but two electrons form closed singlet shells. This allows to obtain the properties of these structures by configuration interaction calculation where the number of electrons is limited to two and where $N-2$ lowest in energy single-electron orbitals forming singlet shells are excluded. Despite the fact that for $N>2$ the properties of the system are controlled by only two electrons the dots with such occupation cannot be treated as an exact equivalent of two-electron systems. We found that the occupation of excited single-electron orbitals by the valence electrons results in different effective $g$-factors in the adjacent dots. For $N>2$ the difference is obtained already for a symmetric system while for two-electrons it results from the dots asymmetry. The differences of effective $g$-factors present in our results are observed in recent EDSR experimental studies on double-quantum dots.

\section{Acknowledgements}
This work was supported by the funds of Ministry of Science and Higher Education for 2014 and by PL-Grid Infrastructure.


\begin{thebibliography}{00}
\bibitem{comment} We name the states according to their spin configuration obtained in the absence of spin-orbit interaction, i.e. as the spins were completely polarized.
\end{thebibliography}
\end{document}